%% file: main.tex
\documentclass[aps,prl,twocolumn,superscriptaddress,groupedaddress]{revtex4-1}

\usepackage{graphicx}   
\usepackage{amsmath}    
\usepackage{amssymb}    
\usepackage{bm}         
\usepackage{hyperref}   

\input{macros.tex}

\input{results_macros.tex}

\begin{document}

\title{A precision measurement of the electron's electric dipole moment using trapped molecular ions}

\author{William B.~Cairncross}
\email{william.cairncross@colorado.edu}
\author{Daniel N.~Gresh}
\author{Matt Grau} \altaffiliation[Present address: ]{Institute for Quantum Electronics, ETH Z\"urich, Otto-Stern-Weg 1, 8093 Z\"urich, Switzerland}
\author{Kevin C.~Cossel} \altaffiliation[Present address: ]{National Institute of Standards and Technology, 325 Broadway, Boulder, Colorado 80305, USA}
\author{Tanya S.~Roussy}
\author{Yiqi Ni} \altaffiliation[Present address: ]{MIT-Harvard Center for Ultracold Atoms, Research Laboratory of Electronics, and Department of Physics, Massachusetts Institute of Technology, Cambridge, Massachusetts 02139, USA}
\author{Yan Zhou}
\author{Jun Ye}
\author{Eric A.~Cornell}
\affiliation{JILA, NIST and University of Colorado, Boulder, Colorado 80309-0440, USA}
\affiliation{Department of Physics, University of Colorado, Boulder, Colorado 80309-0440, USA}

\date{\today}

\begin{abstract}
We describe the first precision measurement of the electron's electric dipole moment (eEDM, $d_e$) using trapped molecular ions, demonstrating the application of spin interrogation times over $700\,{\rm ms}$ to achieve high sensitivity and stringent rejection of systematic errors. Through electron spin resonance spectroscopy on $^{180}$Hf$^{19}$F$^+$ in its metastable $\td$ electronic state, we obtain $d_e = (\deecm \pm \sstatecm_{\rm stat} \pm \ssystecm_{\rm syst})\times 10^{-29}\,\ecm$, resulting in an upper bound of $|d_e| < \ubecm \times 10^{-28}\,\ecm$ (90\% confidence). Our result provides independent confirmation of the current upper bound of $|d_e| < 9.3 \times 10^{-29}\,\ecm$ [J.~Baron {\it et al.}, Science {\bf 343}, 269 (2014)], and offers the potential to improve on this limit in the near future.
\end{abstract}

\maketitle

A search for a nonzero permanent electric dipole moment of the electron (eEDM, $d_e$) constitutes a nearly background-free test for physics beyond the Standard Model (SM), since the SM predicts $|d_e|\lesssim 10^{-38}\,\ecm$ \cite{Pospelov:1991tg}, while the natural scale of $d_e$ in many proposed SM extensions is typically $10^{-27}$ to $10^{-30}\,\ecm$ \cite{Pospelov:2005fw}. Present experimental techniques now constrain these theories \cite{Baron:2014ds}; hence, there have been many recent experimental efforts to measure an eEDM \cite{Baron:2014ds, Hudson:2011hs, Regan:2002ka, Eckel:2013go,  Zhu:2013ct, Heidenreich:2005ka, Lee:2013kn}. 

The most precise eEDM measurements to date were performed using thermal beams of neutral atoms or molecules \cite{Regan:2002ka, Hudson:2011hs, Baron:2014ds}. These experiments benefited from excellent statistical sensitivity provided by a high flux of neutral atoms or molecules, and decades of past work have produced a thorough understanding of their common sources of systematic error. Nonetheless, a crucial systematics check can be provided by independent measurements conducted using different physical systems and experimental techniques. Moreover, techniques that allow longer interrogation times offer significant potential for sensitivity improvements in eEDM searches and other tests of fundamental physics \cite{Skripnikov:2017tu}.

In this Letter, we report on a precision measurement of the eEDM using molecular ions confined in a radio frequency (RF) trap, applying the methods proposed in Ref.~\cite{Leanhardt:2011jv} and demonstrated in Ref.~\cite{Loh:2013dc}. We perform an electron spin precession experiment on $^{180}{\rm Hf}^{19}{\rm F}^+$ molecules in their metastable $\td$ electronic state, and extract the relativistically enhanced eEDM-induced energy shift $\sim 2 d_e {\mathcal E}_{\rm eff}$ between stretched Zeeman sublevels, where $\Eeff \approx 23$ GV/cm in HfF$^+$ \cite{Eeff_footnote,Meyer:2006ky,Petrov:2007hr,Fleig:2013ij,skripnikov_private_comm}. In addition to leveraging the high eEDM sensitivity and systematic error rejection intrinsic to an $|\Omega|=1$ electronic state in a heavy polar molecule \cite{Eckel:2013go}, including in particular the small magnetic moment of a $\td$ state \cite{Meyer:2006ky}, we use a unique experimental approach that is robust against sources of systematic error common to other methods. The 2.1(1)~s lifetime of the $\td$ state in HfF$^+$ \cite{Ni:2014ck} and our use of an RF trap allow us to attain spin precession times in excess of 700~ms -- nearly three orders of magnitude longer than in contemporary neutral beam experiments. This exceptionally long interrogation time allows us to obtain high eEDM sensitivity despite our lower count rate. In addition, performing an experiment on trapped particles permits the measurement of spin precession fringes at arbitrary free-evolution times, making our experiment relatively immune to systematic errors due to initial phase shifts associated with imperfectly characterized state preparation.

\begin{figure}[tb]
        \centering
        \includegraphics[width=\columnwidth]{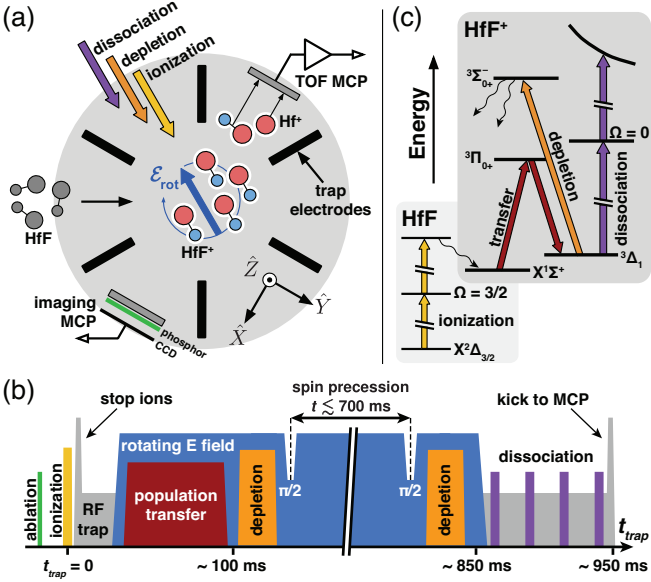}
        \caption{(a) Apparatus schematic, (b) experimental timing, and (c) relevant energy levels (not to scale) for an eEDM measurement using trapped ions. HfF is resonantly photoionized (yellow) to form HfF$^+$. A rotating electric bias field $\Erot$ (blue) polarizes the molecules, and transfer [red, not shown in (a)] and depletion lasers (orange) perform state preparation. The spin resonance sequence is performed by modulating the value of $\Erot$. Spin state populations are detected by depletion followed by resonant multiphoton photodissociation (purple) and counting the resulting Hf$^+$ ions on a time-of-flight microchannel plate detector (TOF MCP).}
        \label{fig:apparatus_timing}
    \end{figure}

Our apparatus and experimental sequence, shown schematically in Fig.~\ref{fig:apparatus_timing}, have been described in detail previously \cite{Leanhardt:2011jv,Loh:2011dn,Loh:2012kp,Loh:2013dc,Ni:2014ck,Cossel:2012hna}. We produce HfF by ablation of Hf metal into a pulsed supersonic expansion of Ar and SF$_6$. The reaction of Hf with SF$_6$ produces HfF, which is entrained in the supersonic expansion and rovibrationally cooled through collisions with Ar. The resulting beam enters the RF trap, where HfF is ionized with pulsed UV lasers at 309.4~nm and 367.7~nm to form HfF$^+$ in its $^1\Sigma^+$, $v=0$ ground vibronic state \cite{Loh:2011dn,Loh:2012kp}. The ions are stopped at the center of the RF trap by a pulsed voltage on the radial trap electrodes, then confined by a DC axial electric quadrupole field and an RF radial electric quadrupole field with frequency $\frf = 50$~kHz. We next adiabatically turn on a spatially uniform electric bias field $\Erot \approx 24$~V/cm that rotates in the radial plane of the ion trap with typical frequency $\frot \approx 250\,{\rm kHz}$, causing the ions to undergo circular motion with radius $r_{\rm rot} \approx 0.5\,{\rm mm}$. A pair of magnet coils aligned with the $Z$ axis produce an axial magnetic gradient $\bm{\mathcal B}=\Baxgrad (2\bm{Z}-\bm{X}-\bm{Y})$ where $|\Baxgrad|\approx 40\,{\rm mG/cm}$, which in the rotating, translating frame of the ions creates a magnetic bias field $\Brot \equiv |\langle \bm{\mathcal B} \cdot \bm{\mathcal E}_{\rm rot}/\Erot \rangle | \simeq |\Baxgrad r_{\rm rot}|$ that is parallel (antiparallel) to $\Erot$ if $\Baxgrad>0$ ($<0$) \cite{Leanhardt:2011jv,Loh:2013dc}. 

\begin{figure}[tb]
        \centering
        \includegraphics[width=\columnwidth]{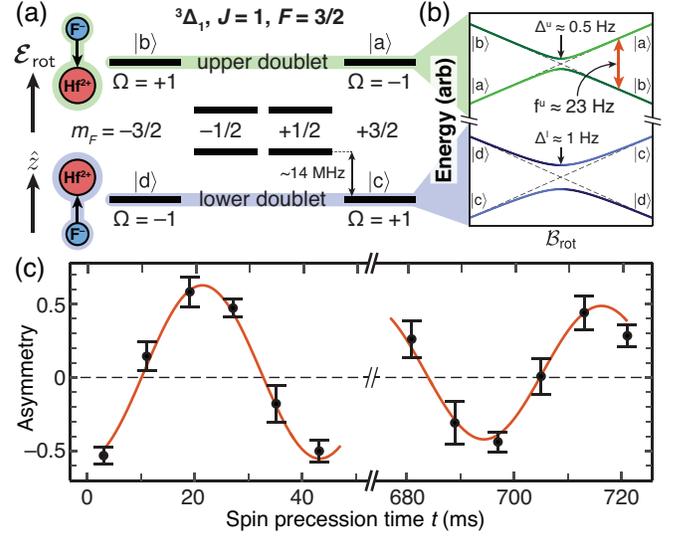}
        \caption{Electron spin resonance spectroscopy in HfF$^+$. (a) Level structure of the eEDM-sensitive $\td$, $J=1$, $F=3/2$ state in an electric bias field $\Erot$. (b) Energies of $|m_F|=3/2$ states as a function of magnetic bias field $\Brot$ (not to scale), showing an avoided crossing at $\Brot=0$ due to a rotation-induced fourth-order coupling $\Delta^{u/l}$ \cite{supplement}. (c) Sample interference fringe with frequency $f^u(\Brot) \approx 23\,\Hz$ indicated in (b), showing an interrogation time of $\sim700\,{\rm ms}$ and decoherence rate $\gamma=0.3(2)\,{\rm s}^{-1}$.}
        \label{fig:precision_spectroscopy}
    \end{figure}

Our state preparation consists of population transfer to the eEDM-sensitive $\td$ state and selective depletion of magnetic sublevels to produce a pure spin state [Fig.~\ref{fig:apparatus_timing}(b-c)]. Two cw lasers at $899.7\,{\rm nm}$ and $986.4\,{\rm nm}$ co-propagating along the $-\hat{Z}$ axis drive a stimulated Raman transition through a $^3\Pi_{0^+}$, $v=1$, $J=1$ intermediate state, transferring approximately $40\%$ of the ground rovibronic state population to the $\td$, $J=1$, $F=3/2$ state. Figure \ref{fig:precision_spectroscopy}(a) shows the structure of this state in a frame defined by the instantaneous direction of $\vecErot \equiv \Erot \hat{z}$. It consists of four Stark doublets (pairs of magnetic sublevels) separated by $\dmf \Erot/3h \approx 14~\MHz$, where $\dmf$ is the $\td$ molecule-frame dipole moment and $h$ is Planck's constant. The population transfer process resolves Stark doublets, but produces an incoherent mixture of $m_F=\pm 3/2$ states in either the upper or lower doublet, depending on the detuning of the second transfer laser. Selective depletion is then performed by a circularly polarized 814.5~nm Ti:sapphire laser resonant with the $P(1)$ line of a $^{3}\Sigma^{-}_{0^+}\leftarrow\td$ transition. The depletion laser is strobed synchronously with the rotating electric field so that its wavevector is either parallel or antiparallel to $\vecErot$, thus driving a $\sigma^\pm$ transition to an $F'=1/2$ manifold and leaving a single $m_F=\pm 3/2$ level populated in the $\td$ state.

Following the production of a pure spin state by strobed depletion, we perform a $\pi/2$ pulse to prepare an equal superposition of $m_F=\pm3/2$ states. This is accomplished by reducing $\Erot$ for a brief interval, which increases a rotation-induced fourth-order coupling $\Delta^{u/l}$ between $m_F=\pm3/2$ states [Fig.~\ref{fig:precision_spectroscopy}(b)] and causes a pure spin state to evolve into an equal superposition in $\sim 1\,{\rm ms}$ \cite{Leanhardt:2011jv,Loh:2013dc,supplement}. We return $\Erot$ to its nominal value and allow the phase of the superposition state to evolve for a variable precession time up to $\sim 700$~ms, then apply a second $\pi/2$ pulse to map the relative phase of the superposition onto a population difference between $m_F=\pm3/2$ states. A second set of strobed laser pulses again depletes all but a single $m_F=\pm 3/2$ level. Finally, to selectively detect the remaining population in the $\td$, $J=1$ state, we resonantly photodissociate HfF$^+$ using pulsed UV lasers at 285.7~nm and 266~nm. We eject all ions from the trap with a pulsed voltage on the radial trap electrodes, and count both Hf$^+$ and the temporally resolved background HfF$^+$ using a microchannel plate (MCP) detector \cite{Ni:2014ck}. 

We interleave experimental trials where the two sets of strobed depletion pulses have the same or opposite phase with respect to $\Erot$ in order to alternately prepare and detect population in the $m_F=\pm 3/2$ states. Denoting by $N_A$ ($N_B$) the measured population when the depletion phases are the same (opposite), we form the asymmetry ${\mathcal A}=(N_A-N_B)/(N_A+N_B)$, which normalizes drifts in the absolute $\td$ population. The asymmetry forms an interference fringe that is well-approximated by a sinusoidal function of precession time $t$,
    \begin{equation}
    \label{eq:asymmetry}
        \mathcal{A}(t) \simeq - \mathcal{C} e^{-\gamma t} \cos(2\pi f t + \phi) + \mathcal{O},
    \end{equation}
with frequency $f$ proportional to the energy difference between the $m_F=\pm 3/2$ states, as shown in Fig.~\ref{fig:precision_spectroscopy}(c). The initial contrast $\mathcal{C}$, initial phase $\phi$, offset $\mathcal{O}$, and decoherence rate $\gamma$ parametrize imperfect state preparation and the loss of coherence, which is primarily due to ion-ion collisions. We perform nonlinear least squares fitting of the asymmetry with the functional form of Eq.~(\ref{eq:asymmetry}) with $\mathcal{C}$, $\gamma$, $f$, $\phi$, and $\mathcal{O}$ as fit parameters. Standard errors $\delta \mathcal{C}$, $\delta \gamma$, $\delta f$, $\delta \phi$, and $\delta \mathcal{O}$ are estimated from the Jacobian of the fit function at the optimum parameter values. The precession frequency contains the eEDM signal, while the other fit parameters are used to diagnose experimental imperfections and sources of systematic error. 

To isolate an eEDM-dependent frequency shift and diagnose systematic errors, we form data ``channels'': components of a measurement that have a particular parity under a set of chosen ``switches'' -- experimental parameters that are modulated between opposite values on a short timescale \footnote{Our notation for data channels closely follows that used by the authors of Refs.~\cite{Baron:2014ds,Baron:2016vr}.}. Our switches are the sign of the magnetic bias field $\tilB = {\rm sgn}(\langle \bm{\mathcal B} \cdot \vecErot \rangle)$, the populated Stark doublet $\tilD = -{\rm sgn}(m_F \Omega)$, and the sense of the electric bias field rotation $\tilR = -{\rm sgn}(\bm{\omega}_{\rm rot} \cdot \hat{Z})$. We repeat our spin precession measurement in each of the eight unique ``switch states'' $\tilS=(\tilB,\tilD,\tilR)$ to form a ``block,'' and form channels $X^s$ with parities $s \subset \{B,D,R\}$ from linear combinations of the eight measurements $X(\tilS)$, where $X\in \{{\mathcal C}, \gamma, f, \phi, {\mathcal O}\}$ (See Eq.~(S1) in Ref.~\cite{supplement}). For example, for a given block of data, $f^B$ is given by half the difference between the average value of $f$ for the $\tilB=+1$ fringes and that for the $\tilB=-1$ fringes. We estimate the standard error $\delta X$, which is the same for all parities $s$, by propagating the error estimates $\delta X(\tilS)$ resulting from the nonlinear least squares fit of Eq.~(\ref{eq:asymmetry}).

If higher order effects are neglected, the measured spin precession frequency is dominated by the Zeeman shift between populated magnetic sublevels, and includes a $BD$-odd contribution from an eEDM:
    \begin{equation}
    \begin{split}
        h f(\tilS) &\approx \left| - 3 g_F \mu_B \tilB \Brot + 2\tilD d_e |\Eeff| \right| \\
        &= 3 |g_F| \mu_B \Brot - 2 \tilB \tilD \, \sgn(g_F) \, d_e |\Eeff|.
    \end{split}
    \end{equation}
An eEDM signal thus appears as the lowest-order contribution to the $f^{BD}$ frequency channel, while any non-ideal contributions to $f^{BD}$ constitute potential sources of systematic error. The seven non-eEDM frequency channels contain information about experimental conditions such as non-reversing magnetic fields, and we use these channels to construct and confirm models of non-ideal experimental behavior and to correct for systematic shifts in $f^{BD}$. Some examples of frequency channels, their leading-order expressions in terms of experimental parameters, and their physical interpretations are shown in Table~\ref{tab:lcoms}.
    \begin{table}[t]
    \centering
    \begin{tabular}{l  l  r}
        \hline \hline
         Channel
            & Leading term 
                & Interpretation  \\ \hline
         $f^0$
            & $3 |g_F| \mu_B \Brot / h$
                & Avg.~precession frequency \\ 
         $f^B$      
            & $3 |g_F| \mu_B \Brotnr / h$
                & Non-reversing $\Brot$ \\
         $f^D$
            & $3 \dgeff \mu_B \Brot \sgn(g_F) / h$
                & Level-dependent g-factor \\ 
         $f^{BR}$
            & $-3 \langle\alpha\rangle \frot \sgn(g_F)$
                & Geometric phase \\ 
         $f^{BD}$
            & $-2 d_e |\Eeff| \, \sgn(g_F) / h$
                & eEDM shift \\
         \hline \hline
    \end{tabular}
    \caption{Selected frequency channels, their leading expression in terms of experimental parameters, and their interpretations. Here $\dgeff$ is half the effective magnetic g-factor difference between Stark doublets, $\alpha$ is the tilt angle of $\vecErot$ above the radial plane of the ion trap, and $\frot$ is the rotation frequency of $\Erot$.}
    \label{tab:lcoms}
    \end{table}
    \begin{figure}[t]
        \centering
        \includegraphics[width=\columnwidth]{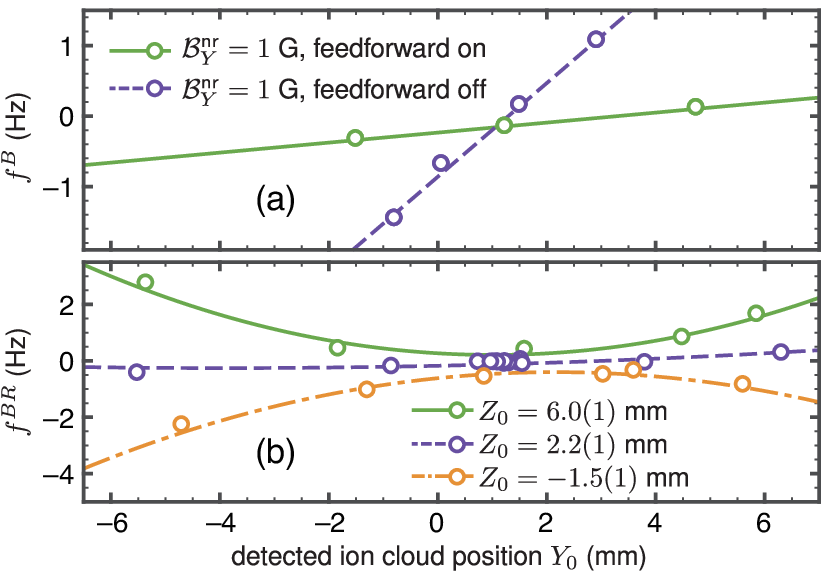}
        \caption{Non-ideal frequency shifts in the $f^{B}$ and $f^{BR}$ channels due to a stray uniform magnetic field $\BYnr$ and ion displacements $Y_0 \hat{Y}$ and $Z_0 \hat{Z}$ \cite{supplement}. (a) A shift in $f^B \propto \BYnr Y_0$ resulted from a contribution to $\Brot$ from an electric field gradient oscillating at $2 \frot$, which we suppressed by reducing harmonic distortion in $\Erot$ via feedforward. (b) A shift in $f^{BR}=3 \langle \alpha \rangle \frot \propto Y_0^2 Z_0$ was well modeled by the known inhomogeneity in $\Erot$, and was suppressed by applying feedback to the ion position between eEDM measurements. Error bars are $\sim \pm 0.1~\Hz$ on all points.}
        \label{fig:systematics}
    \end{figure}
Prior to eEDM data collection, we tuned a wide variety of experimental parameters over an exaggerated dynamic range and observed the response of the data channels in order to create models of non-ideal frequency shifts in our system. Two illustrative examples of these observations are shown in Fig.~\ref{fig:systematics}, and the contributions of these shifts to the eEDM channel are discussed in Ref.~\cite{supplement}. The result of this study was the validation of a unified numerical model of our spin precession sequence. In this model, we integrate the classical motion of ions in simulated time-varying electric and magnetic fields, then propagate the internal quantum state of the molecules using an effective Hamiltonian that explicitly includes all sublevels of the two lowest rotational levels of $\td$. Using known experimental parameters and realistic estimates of construction imperfections, this model was able to reproduce all observed frequency shifts.

\begin{figure}[tb]
        \centering
        \includegraphics[width=\columnwidth]{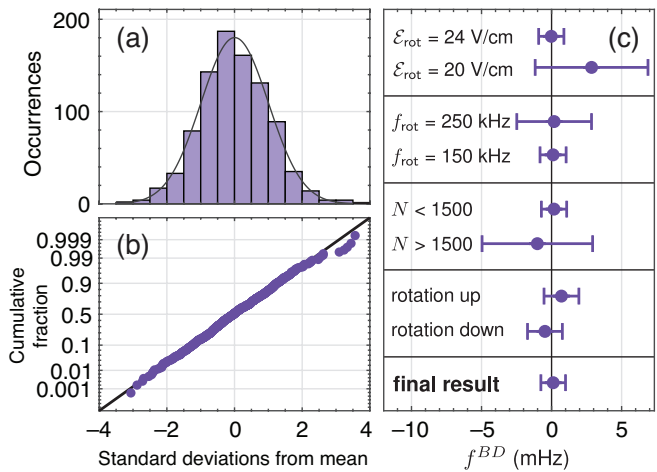}
        \caption{Summary of eEDM dataset after cuts and scaling $\delta f$ by $\sqrt{\chi_r^2}$ to account for over-scatter. (a) Histogram of normalized and centered eEDM-sensitive frequency channel measurements $(f^{BD} - \langle f^{BD} \rangle)/\delta f$. (b) Normal probability plot of the same dataset, showing a linear trend suggesting that the data are consistent with a normal distribution. (c) Subsets of the eEDM data taken under different values of experimental parameters, as well as the overall average of $f^{BD}$. Here $N$ is the average number of trapped HfF$^+$ ions per experimental trial.}
        \label{fig:eedm_data}
    \end{figure}

In total, we collected 1024 blocks (360.3 hours) of eEDM-sensitive data, with each block resulting in one value of $f^{BD}$ and thus one eEDM measurement. Throughout the collection and analysis of this eEDM data, we added to the $f^{BD}$ channel an unknown, computer-generated pseudo-random value drawn from a normal distribution with a standard deviation of $5\times 10^{-28}\,\ecm$. This ``blind'' allowed us to investigate systematic frequency shifts and perform statistical analysis while mitigating the effects of operator bias. We applied cuts to the blinded data based on non-eEDM channels indicating signal quality: blocks with ${\mathcal C} < 0.1$ or ${\mathcal C} e^{-\gamma T} < 0.1$ were cut due to low signal to noise (where $T$ is the largest value of $t$ sampled in a block). In addition, we cut data where shifts in the ``co-magnetometer'' channel $f^B$ exceeded $0.4~\Hz$ due to its contribution to systematic errors. After these cuts, our eEDM dataset consists of 903 blocks or 313.8 hours of data. The unblinded dataset is shown in Fig.~\ref{fig:eedm_data}(a-b). Visual inspection of a normal probability plot, as well as Kolmogorov-Smirnov and Shapiro-Wilk normality tests indicate that the distribution of normalized and centered eEDM measurements $(f^{BD} - \langle f^{BD} \rangle)/\delta f$ is consistent with a normal distribution. The reduced chi-squared statistic for fitting a weighted mean to the eEDM dataset is $\chi^2_r = 1.22(5)$. This over-scatter is attributable to magnetic field drifts \cite{supplement}, and to correct for it we scale our final statistical error bar by $\sqrt{\chi^2_r} \approx 1.1$.

During eEDM data collection, we suppressed sources of systematic error that appeared in our earlier model-building investigation by applying active feedback to relevant experimental parameters between the collection of data blocks. The only one of these that produced an observable shift in the $f^{BD}$ channel was the combined effect of a non-reversing magnetic bias field $\Brotnr$ and the difference in effective magnetic g-factor between Stark doublets, which in our system arises from Stark mixing with $\td$, $J=2$ and from our rotating quantization axis \cite{supplement,Leanhardt:2011jv}. The $f^B$ and $f^D$ frequency channels, which are acquired concurrently with $f^{BD}$, provide direct measurements of these contributions. Since the value of $f^D \approx 10^{-3} f^0$ is fixed by the values of $\Erot$, $\Brot$, and $\frot$, we suppress the systematic shift in the eEDM channel by applying a compensating $\Baxgrad$ to minimize $|f^B|$. We also apply a block-by-block correction to $f^{BD}$ based on the measured values of $f^B$ and $f^D$, the validity of which was verified in our earlier model-building study \cite{supplement}. 

Though they were too small to be observed at our level of sensitivity, we predicted systematic shifts in the eEDM channel due to the non-ideal frequency shifts in the $f^B$ and $f^{BR}$ channels shown in Fig.~\ref{fig:systematics}. We suppressed the first of these by adding a feedforward signal to $\Erot$ to cancel the harmonic distortion component at $2 \frot$, reducing it from $-48$~dBc to $-70$~dBc, and by using magnet coils to null the ambient uniform magnetic field at the RF trap center to within $\sim \pm 30$~mG. To suppress the shift in $f^{BR}$ caused by $\Erot$ inhomogeneity shown in Fig.~\ref{fig:systematics}(b), we measured the ion cloud position once per data block on a pair of MCPs, and applied DC potentials on the trap electrodes to position the ion cloud within $\sim 2\,{\rm mm}$ of the minimum of the quadratic shift. The residual offset of $f^{BR} \approx -100~\mHz$ and gradient of $\partial f^{BR}/\partial Y_0 \approx  20\,{\rm mHz/mm}$ shown in Fig.~\ref{fig:systematics} are consistent with $\Erot$ inhomogeneity resulting from realistic machining, welding, and assembly imperfections in the construction of our RF trap.

While collecting eEDM data, we also searched for new systematic errors correlated with parameters that could not be tuned over a significantly exaggerated dynamic range, including $\Erot$, $\frot$, and the number of HfF$^+$~ions trapped per experimental trial [Fig.~\ref{fig:eedm_data}(c)]. We did not observe significant variation of $f^{BD}$ (or five of seven other non-eEDM frequency channels) with these parameters at our current level of precision. The variations of the non-eEDM frequency channels $f^0$ and $f^D$, in which we did anticipate variation with $\Erot$ and $\frot$, were consistent with model predictions. Finally, we modified our data collection by randomizing the order of switch states in each block to search for and suppress systematic errors caused by parameter drifts correlated with our switches, and observed no significant variation of data channels \cite{supplement}. The final results of our systematic error searches and corrections are summarized in Table~\ref{tab:systematics}. 
    \begin{table}[htb]
        \centering
        \begin{tabular}{l c c}
            \hline \hline
            Effect 
                & Correction 
                    & Uncertainty 
                        \\ \hline
            Non-reversing $\Brot$
                & $\fcorrA$
                    & $\sfcorrA$
                        \\
            Geometric phases
                & 
                    & $\sftotB$
                        \\
            Axial secular motion
                & 
                    & $\sftotC$
                        \\ 
            Rotation-odd $\Erot$
                & 
                    & $\sftotE$
                        \\
            Doublet population background
                & 
                    & $\sftotG$ \\
            \hline
            \hfill Total systematic
                & $\fcorrA$
                    & $\ssystuHz$
                        \\
            \hfill Statistical
                & 
                    & $\sstatuHz$
                        \\
            \hfill Total uncertainty
                & 
                    & $\stotuHz$
                        \\
            \hline \hline
        \end{tabular}
        \caption{Systematic effects and corrections applied to the eEDM channel $f^{BD}$, in units of $\uHz$ (with $1\,\uHz$ corresponding to $\sim 10^{-31}\,\ecm$) \cite{supplement}.}
        \label{tab:systematics}
    \end{table}

We removed our blind on 31~March~2017, and obtained a final value for the eEDM-sensitive frequency channel 
    \begin{equation}
        f^{BD} = \feedmmHz \pm \sstatmHz_{\rm stat} \pm \ssystmHz_{\rm syst} ~ {\rm mHz}.
    \end{equation}
Dividing by $-2 |\Eeff| \sgn(g_F)/h \approx \mhztoecm \times 10^{28}~\mHz/\ecm$ \cite{Petrov:2007hr,Fleig:2013ij}, we obtain a value for the eEDM
    \begin{equation}
        d_e = (\deecm \pm \sstatecm_{\rm stat} \pm \ssystecm_{\rm syst}) \times 10^{-29} ~ \ecm,   
    \end{equation}
which is consistent with zero within one standard error. The resulting upper bound is
    \begin{equation}
        |d_e| < \ubecm \times 10^{-28} \, \ecm \quad \text{(90\% confidence)}.
    \end{equation}
Our result is consistent with the limit of $|d_e| < 9.3\times 10^{-29}~\ecm$ set by the ACME Collaboration \cite{Baron:2014ds,Baron:2016vr}, and we have confirmed their result using a radically different experimental approach. Our measurement is limited by statistics, and our dominant source of systematic error can be further suppressed to the $10^{-30}\,\ecm$ level \cite{supplement}. Here we have assumed that parity and time-reversal violating effects arise purely from $d_e$. An additional contribution $\sim W_S C_S$ can arise from a pseudoscalar-scalar electron-nucleon coupling $C_S$ \cite{Chupp:2015hh,Skripnikov:2016bw,Denis:2015ez}, however to our knowledge a published value of $W_S$ is not yet available for HfF$^+$. 

Since the completion of this first generation eEDM measurement, we have designed and constructed a second generation ion trap, which we will use to confine up to an order of magnitude more ions, cool them via adiabatic expansion to a volume up to one hundred times larger, and polarize them using a rotating electric bias field that is more uniform due to an improved electrode design. We estimate that these and other improvements should provide an order of magnitude higher eEDM sensitivity. In the further future, we plan to pursue a third generation eEDM measurement using $^{232}$Th$^{19}$F$^+$, in which the $\td$ ground electronic state with $\Eeff\approx 36\,{\rm GV/cm}$ may allow a high-sensitivity eEDM experiment with a coherence time up to tens of seconds \cite{Gresh:2016hh,Skripnikov:2015fs,Denis:2015ez}.

\begin{acknowledgments}
We thank F.~Abbasi-Razgaleh for experimental assistance, and J.~Bohn, H.~Lewandowski, K.B.~Ng, B.~Spaun, and J.~Thompson for discussions. W.B.C.~acknowledges support from the Natural Sciences and Engineering Research Council of Canada. This work was supported by the Marsico Foundation, NIST, and the NSF (Award PHY-1125844).
\end{acknowledgments}

\bibliographystyle{apsrev4-1} 
\bibliography{main.bbl}

\end{document}


\title{Supplemental material for ``A precision measurement of the electron's electric dipole moment using trapped molecular ions''}

\author{William B.~Cairncross}
\author{Daniel N.~Gresh}
\author{Matt Grau}
\author{Kevin C.~Cossel}
\author{Tanya S.~Roussy}
\author{Yiqi Ni}
\author{Yan Zhou}
\author{Jun Ye}
\author{Eric A.~Cornell}

\affiliation{JILA, NIST and University of Colorado, Boulder, Colorado 80309-0440, USA}
\affiliation{Department of Physics, University of Colorado, Boulder, Colorado 80309-0440, USA}

\maketitle

\section{Data collection} \label{sec:data_collection}

\subsection{Switch state timing \& control}
As described in the main text, a single eEDM measurement requires collecting one interference fringe in each of the eight unique switch states $(\tilB,\tilD,\tilR)$. As a compromise between signal to noise on a single fringe and reducing susceptibility to errors from drifting experimental conditions, we typically collect twelve points per fringe, with six points spanning one fringe period at short spin precession times $0<t<40~{\rm ms}$ and six more points spanning a second fringe period at long spin precession time, $(T-40~{\rm ms}) < t < T$, where during data collection we varied $T$ between 200 and 700 ms. One measurement of the asymmetry ${\mathcal A}(t)$ requires two experimental trials, and we typically average eight measurements comprising sixteen trials at each value of $t$. As shown in Fig.~1(b) in the main text, a single experimental trial is accompanied by approximately $300~{\rm ms}$ of ``dead time'' spent on state preparation, state readout, auxiliary measurements, and saving data. Thus, a single block requires a minimum of approximately 16 minutes to complete. 

\begin{figure}[h]
        \centering
        \includegraphics[width=0.8\textwidth]{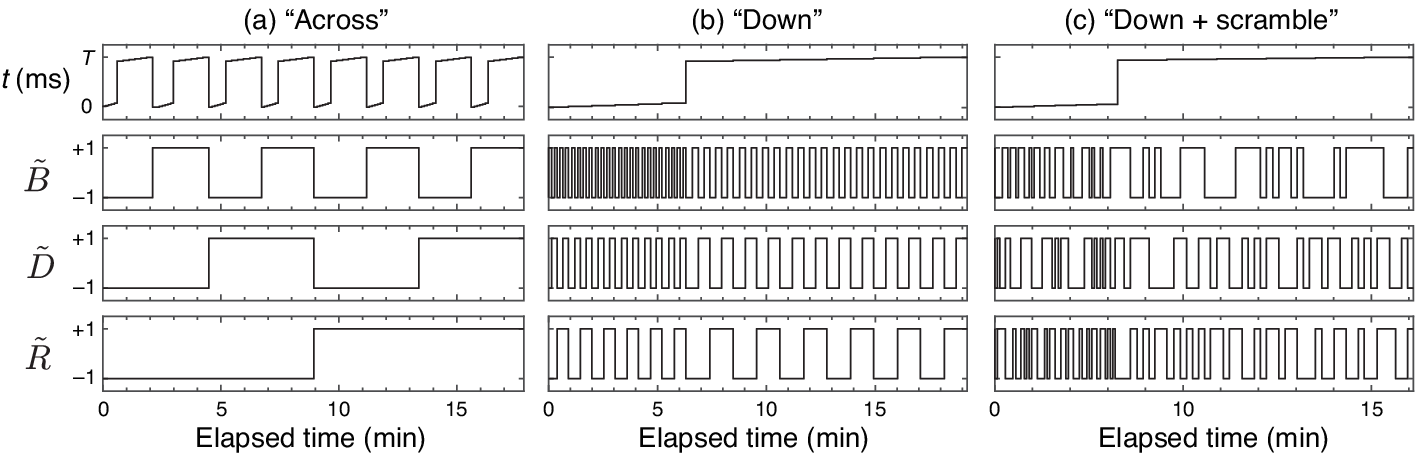}
        \caption{Relative timing of experimental configurations during the collection of one data block. (a) When taking data ``across,'' we collect in each switch state an entire interference fringe as a function of $t$. (b) In the ``down'' configuration, we change the spin precession time $t$ on a slower timescale than $B$, $D$, and $R$. (c) In the ``down + scramble'' configuration, we also re-randomize the order of switch states $(\tilB,\tilD,\tilR)$ at each value of $t$.}
        \label{fig:switch_timing}
    \end{figure}

The timescale of data collection and the order of switch states and interrogation times can affect our statistical sensitivity, susceptibility to parameter drift, and sources of systematic error, depending primarily on the Fourier spectrum of ${\mathcal B}$ field drifts. We used three different ordering schemes, shown in Fig.~\ref{fig:switch_timing}, to investigate and mitigate these errors. The first, ``across,'' makes up the smallest fraction of our eEDM dataset. We expect the ``down'' timing scheme to be less susceptible to drifts in experimental parameters on the few-minute timescale, for example due to temperature fluctuations or the accumulation of patch potentials on ion trap surfaces. The ``down + scramble'' timing scheme, in which the order of switch states is randomized in each block, should be still less susceptible to errors associated with both parameter drifts and systematic errors associated with the order of switch states. 

\begin{figure}[h]
        \centering
        \includegraphics[width=0.6\textwidth]{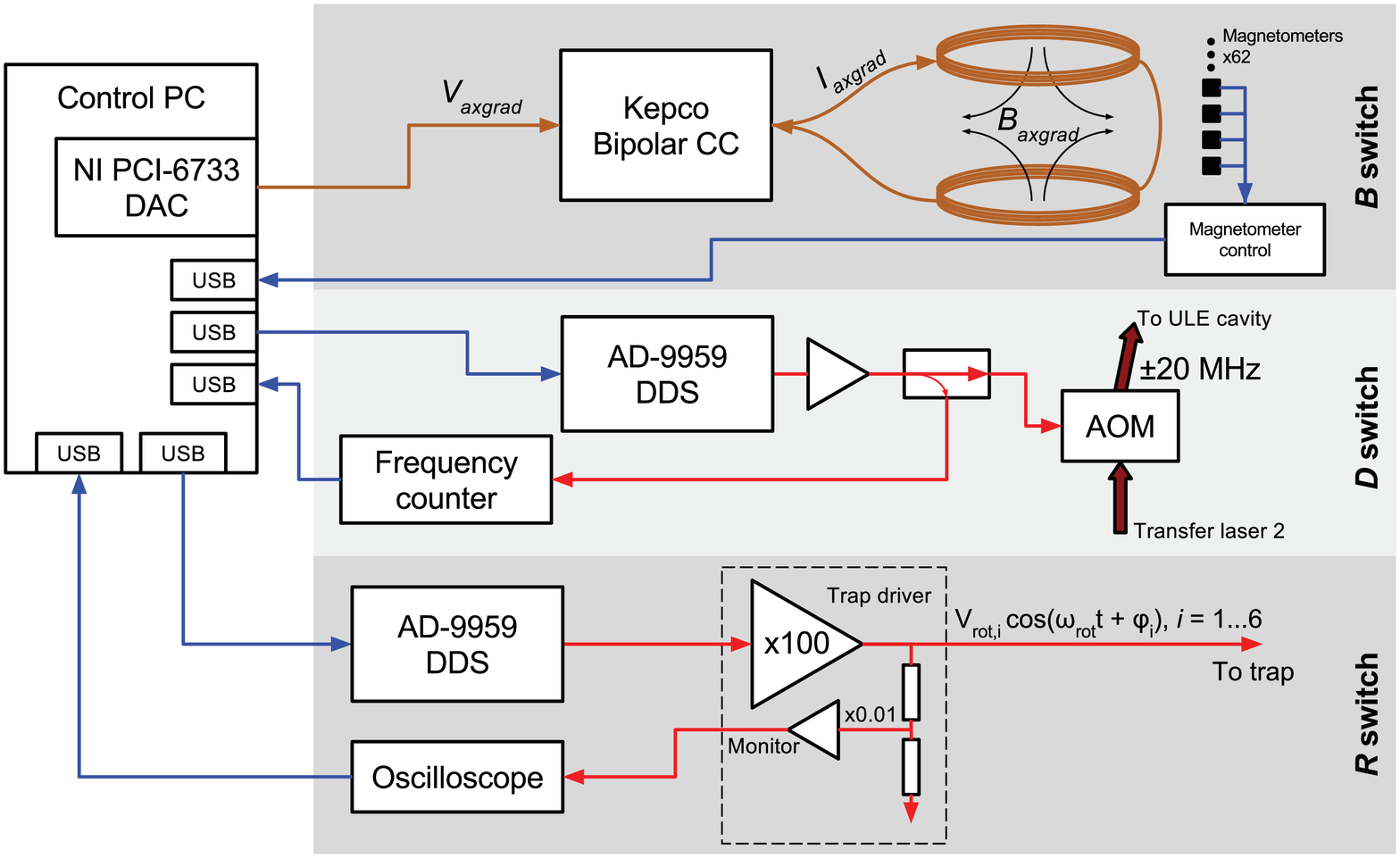}
        \caption{Schematic layout of control system for experimental switches. Blue traces indicate digital signals, gold traces indicated DC analog signals, and red traces indicate RF analog signals. }
        \label{fig:switch_control}
    \end{figure}

As discussed in the main text, the $B$, $D$, and $R$ switches represent the sign of the axial magnetic gradient $\Baxgrad$ generating the rotating magnetic bias field, the populated Stark doublet, and the sense of electric bias field rotation, respectively. A National Instruments PCI-6733 digital-to-analog converter (DAC) supplies a control voltage to a bipolar current supply that powers the pair of magnet coils generating $\Baxgrad$, thus setting the value of $\tilB$. The doublet switch $\tilD$ is set by adjusting the frequency output of an Analog Devices AD9959 direct digital synthesis (DDS) ASIC between two values separated by $\sim20~\MHz$. The amplified output drives an acousto-optic modulator that controls the frequency of the second of our Raman transfer lasers via a frequency offset lock to a stable optical cavity, thus tuning the laser to populate either the upper or lower Stark doublet. Finally, the value of $\tilR$ is set by adjusting the relative phase of six DDS-generated sinusoidal signals that are amplified to produce the rotating electric bias field. These control systems are shown schematically in Fig.~\ref{fig:switch_control}. None of our switches generate or require large currents or voltages, and each can be changed on a timescale that is short compared to collecting one 16-shot data point. Thus we are not restricted in their order due to dead-time considerations, and can randomize our switches without significant change in our duty cycle. 

\section{Data processing}
We use MATLAB to perform data analysis, which consists of counting Hf$^+$ ions at each time point in the fringe, calculating and fitting the asymmetry, forming data channels, blinding the eEDM channel, applying cuts, and searching for signs of systematic errors in the resulting data channels.

\begin{figure}[h]
        \centering
        \includegraphics{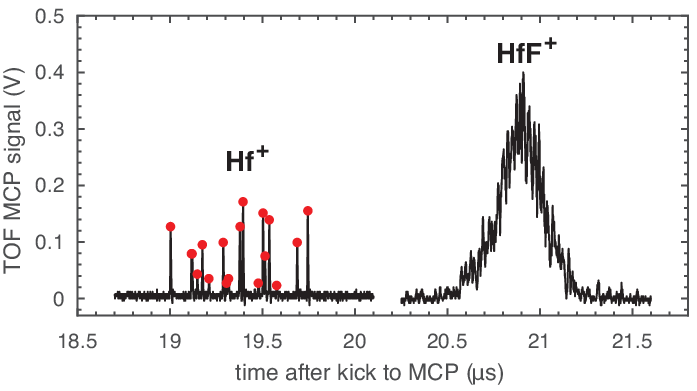}
        \caption{Typical raw data trace from the time-of-flight microchannel plate detector, showing time-resolved Hf$^+$ and HfF$^+$ signals. Red markers indicate counted Hf$^+$ ions contributing to the asymmetry signal.}
        \label{fig:tof_mcp}
    \end{figure}

A typical signal from our time-of-flight (TOF) microchannel plate (MCP) ion detector is shown in Fig.~\ref{fig:tof_mcp}. The ``signal'' Hf$^+$ ions resulting from state-selective photodissociation and the ``spectator'' HfF$^+$ ions are temporally separated, and for numbers of Hf$^+$ ions below $\sim 30$, individual Hf$^+$ ion peaks are also well-resolved. We separately amplify the Hf$^+$ and HfF$^+$ signals to reduce noise on the Hf$^+$ signal while avoiding saturation of our transimpedance amplifier by the HfF$^+$ signal. We use a peak-finding algorithm to locate Hf$^+$ peaks of a specified prominence above the background value; typically 4 times the rms voltage of an empty trace. 

We compute the asymmetry ${\mathcal A}(t)$ from the number of counted Hf$^+$ ions in runs of the experiment with opposite depletion phases. We fit the functional form of Eq.~(1) to the measured asymmetry using MATLAB's Levenberg-Marquardt nonlinear least squares regression algorithm. Because measurements of the asymmetry do not follow a normal distribution (see e.g.~Ref.~\cite{Hutzler:2014wf}), we perform an unweighted regression, rather than binning measurements of ${\mathcal A}(t)$ for a given $t$ and performing a weighted regression. Fitting simulated data with our analysis routine does not show any evidence of systematic shifts due to the non-gaussian distribution of ${\mathcal A}(t)$.

As discussed in the main text in the paragraph preceding Eq.~(2), we form linear combinations of frequency measurements $f^{\tilB \tilD \tilR} \equiv f(\tilB,\tilD,\tilR)$ to obtain frequency channels according to the transformation
    \begin{equation} \label{eq:lcoms}
        \begin{pmatrix}
            f^0 \\
            f^B \\ 
            f^D \\
            f^{BD} \\
            f^R \\
            f^{BR} \\
            f^{DR} \\
            f^{BDR}
        \end{pmatrix} = \frac{1}{8}
        \begin{pmatrix}
            + & + & + & + & + & + & + & + \\
            + & - & + & - & + & - & + & - \\
            + & + & - & - & + & + & - & - \\
            + & - & - & + & + & - & - & + \\
            + & + & + & + & - & - & - & - \\
            + & - & + & - & - & + & - & + \\
            + & + & - & - & - & - & + & + \\
            + & - & - & + & - & + & + & - \\
        \end{pmatrix}
        \begin{pmatrix}
            f^{+++} \\
            f^{-++} \\
            f^{+-+} \\
            f^{--+} \\
            f^{++-} \\
            f^{-+-} \\
            f^{+--} \\
            f^{---} 
        \end{pmatrix},
    \end{equation}
where $\pm$ signs represent $\pm 1$. The standard error $\delta f$ is equal for all frequency channels within one block, and is given by 
    \begin{equation}
        \delta f = \frac{1}{8} \sqrt{\sum_{\tilB,\tilD,\tilR = \pm} (\delta f^{\tilB\tilD\tilR})^2},
    \end{equation}
where $\delta f^{\tilB\tilD\tilR}$ are the standard error estimates obtained from nonlinear least squares regression. We follow the same prescription to obtain the data channels for contrast, phase, decoherence rate, and offset parameters. Within the same MATLAB script that computes this linear transformation of measurements from the ``state basis'' to the ``parity basis,'' we apply the blinding value to the $f^{BD}$ channel. The pseudo-random blind value, which was previously generated and saved to a binary file, is read and added to the $f^{BD}$ channel within a single line of code. Following systematics corrections (discussed in Section~\ref{sec:systematics}), we obtain the eEDM result by computing the weighted mean value of $f^{BD}$ across all blocks (indexed by $n$),
    \begin{equation} \label{eq:weighted_mean}
        \langle f^{BD} \rangle = \frac{\sum_n W_n f^{BD}_n}{\sum_n W_n}, \qquad
        \langle \delta f^{BD} \rangle = \frac{1}{\sqrt{\sum_n W_n}}
    \end{equation}
where we use weights $W_n = (\delta f_n)^{-2}$. To correct for the over-scatter of our data, we scale our final value of $\langle \delta f^{BD} \rangle$ by $\sqrt{\chi^2_r}$, where $\chi^2_r$ is the reduced chi squared statistic of the distribution of normalized and centered eEDM measurements $(f^{BD} - \langle f^{BD} \rangle)/\delta f$. This over-dispersion appears in our frequency channels in a way that is consistent with it arising entirely from a drifting ambient magnetic field gradient $\Baxgradnr$. 

We chose our data cuts based on signal-to-noise considerations and the values of eEDM-insensitive data channels. The first data cut is performed at the ion counting level by a choice of a time-of-flight window and minimum pulse height for Hf$^+$ ion counting. Further cuts were based on initial and final contrast and on the value of the $f^B$ data channel. We investigated the values of frequency channels as a function of the cut parameters, and found no worrisome dependencies.

\section{Modeling frequency channels}
The high statistical sensitivity and the systematic error rejection features of our experimental approach come with an associated cost in the form of an increased level of complexity in modeling frequency measurements. The $\Omega$-doublet structure of the $\td$ state, the nuclear hyperfine structure of HfF$^+$, our rotating quantization axis, and the motion of ions in inhomogeneous and time-dependent electric and magnetic fields all contribute to this complexity. As a result, we used a variety of numerical and perturbative techniques to analyze sources of non-ideal frequency shifts in our system. In this section we discuss our methods in general terms, and provide a list of observed frequency shifts. In Section \ref{sec:systematics} we describe in more detail effects that systematically affect the eEDM measurement channel $f^{BD}$.

\subsection{Ion motion} \label{sec:ion_motion}
The spin precession frequency of HfF$^+$ in our experiment is set by the electric and magnetic fields experienced by each ion. Since the electric and magnetic fields in the RF trap are inhomogeneous and (for the electric fields) time-varying, and the ions' trajectories are modified by the electric field (neglecting the Lorentz force), we must know the trajectories of the ions in order to determine their spin precession frequencies. 

For our perturbative analysis of frequency channels, we use the standard approximation of harmonic motion with superimposed RF micromotion \cite{Berkeland:1998cw}, and additionally superimpose rotating micromotion due to $\Erot$. In this case, the total electric field is
\begin{equation}
    \bm{\mathcal E}(\bm{R},t) = \frac{V_{\rm rf}}{R_0^2} \cos{(\wrf t)} (\bm{X} - \bm{Y}) 
        + \frac{V_{\rm dc}}{Z_0^2} (\bm{X} + \bm{Y} - 2 \bm{Z})
        + \Erot \left[ \hat{X} \cos{(\wrot t)} - \tilR \hat{Y} \sin{(\wrot t)} \right],
\end{equation}
where $\wrot\equiv 2 \pi \frot$, $\wrf\equiv 2 \pi \frf \approx 50~\kHz$, $\tilR$ is the rotation switch sign, and $R_0$ and $Z_0$ are the effective radius and height of the RF trap. However, this approximation is not able to account for effects we observe due to electric field inhomogeneities of multipole order $l>2$. We account for these higher order effects numerically by performing a multipole fit up to $l=9$ of the electric field due to unit potential on each of the eight trap electrodes to obtain multipole coefficients $c_{lm}^k$ (where $k=1\dots8$), which allows us to represent the total electric field in the ion trap as a function of the electrode voltages:
    \begin{equation} \label{eq:Etot}
        \bm{\mathcal E}(\bm{R},t) = \sum_{k=1}^8 V_k(t) \sum_{lm} c_{lm}^k \left[ -\bm\nabla \left( R^l \, Y_{lm}(\Theta,\Phi)\right) \right],
    \end{equation}
where $(R,\Theta,\Phi)$ are spherical polar coordinates in the laboratory frame, and $Y_{lm}$ are real spherical harmonics. We then numerically integrate the equations of motion for an ion to obtain $\bm{R}_{\rm ion}(t)$, and substitute back into Eq.~(\ref{eq:Etot}) and a corresponding expression for $\bm{\mathcal B}(\bm{R})$ to obtain $\bm{\mathcal E}_{\rm ion}(t)$ and $\bm{\mathcal B}_{\rm ion}(t)$, the electric and magnetic fields at the location of the ion. Finally, we transform these fields into a rotating frame whose coordinate axes $\hat{x}$, $\hat{y}$, $\hat{z}$ are related to the laboratory frame axes $\hat{X}$, $\hat{Y}$, $\hat{Z}$ by
    \begin{equation} \label{eq:coords}
        \hat{x} = -\hat{Z}, \qquad
        \hat{y} = \hat{Y} \cos{(\wrot t)} + \tilR \hat{X} \sin{(\wrot t)}, \qquad
        \hat{z} = \hat{X} \cos{(\wrot t)} - \tilR \hat{Y} \sin{(\wrot t)}.
    \end{equation}
In this frame, $\Erot$ nominally points along the $+\hat{z}$ axis. For this analysis, we neglect the effect of ion-ion interactions.

\subsection{Effective Hamiltonian} \label{sec:Heff}
Our modeling of frequency channels relies on an effective Hamiltonian description of HfF$^+$ molecules. We do not use optical pumping into dark states to perform $\pi/2$ pulses, so high-energy photons are not present, and coupling to other electronic states is absent throughout our spin precession sequence. As a result, we can build an effective Hamiltonian that includes only $\td$ sublevels. The $\td$ state is well-described by Hund's case (a) basis states with coupled nuclear spin,
    \[
        \ket{\Lambda=\pm 2, S=1, \Sigma=\mp 1, J, \Omega=\pm 1, I=1/2, F, m_F},
    \]
where $\Lambda={\bf L} \cdot {\bf n}$ is the projection of the electronic orbital angular momentum ${\bf L}$ on the internuclear axis ${\bf n}$, $S=|{\bf S}|$ is the total electronic spin angular momentum, $\Sigma={\bf S}\cdot{\bf n}$ is the electron spin projection on the internuclear axis, $J=|{\bf J}|=|{\bf L}+{\bf S}+{\bf R}|$ is the electronic plus rotational angular momentum, $\Omega={\bf J} \cdot {\bf n}$ is the projection of the electronic angular momentum onto the internuclear axis, $I=|{\bf I}|$ is the $^{19}$F nuclear spin, $F=|{\bf F}|=|{\bf J} + {\bf I}|$ is the total angular momentum of the molecule, and $m_F = {\bf F} \cdot \hat{z}$ is the projection of ${\bf F}$ on the rotating quantization axis \cite{Brown:2003ub}. We take the internuclear axis $\bf n$ to be directed from the $^{19}$F nucleus to the $^{180}$Hf nucleus. We model our spin precession experiment using an effective Hamiltonian that includes (in decreasing order of size) molecular rotation, the nuclear spin hyperfine interaction, the Stark effect, $\Omega$-doubling, a rotating quantization axis, the electronic and nuclear Zeeman effects, and an eEDM:
    \begin{equation}
        H(\bm{\mathcal E},\bm{\mathcal B},\vecwrot) = 
            \Htum + \Hhf + \HS(\bm{\mathcal E}) + \HOD + \Hrot(\vecwrot)
            + \HZe(\bm{\mathcal B}) + \HZN(\bm{\mathcal B}) + \Hedm.
        \label{eq:Heff}
    \end{equation}
We use the effective operators
    \begin{align*} \label{eq:Heff2}
        & \Htum = B_e {\bf J}^2, 
            &&\Hhf = \Apar ({\bf I} \cdot {\bf n}) ({\bf J} \cdot {\bf n}),
                &&\HS = - \dmf {\bf n} \cdot \bm{\mathcal E},
                    &&\HOD = \hbar \wef {\bf \Omega}_x/2,   \\
        & \Hrot = - \hbar \vecwrot \cdot {\bf F}, 
            &&\HZe = - \Gpar \mu_B ({\bf J} \cdot {\bf n}) (\bm{\mathcal B} \cdot {\bf n}),
                &&\HZN = - g_N \mu_N {\bf I} \cdot \bm{\mathcal B},
                    &&\Hedm = -d_e |\Eeff| \Omega,
    \end{align*}
with constants listed in Table~\ref{tab:constants}. The effective operator $\bm{\Omega}_x$ has matrix elements $\delta_{\eta',\eta}\delta_{\Omega',-\Omega}$ (where $\eta$ represents all other quantum numbers).
    \begin{table}[htb]
        \centering
        \begin{tabular}{l l l r}
        \hline \hline
            Constant \hspace{10pt}  &   Value                                   
                & Description                                   & Reference             \\ \hline
            $B_e/h$                   &   $8.983(1)$ GHz                          
                & Rotational constant                           & \cite{Cossel:2012hna} \\
            $\Apar/h$                 &   $-62.0(2)$ MHz                          
                & Hyperfine constant                            & This work \\
            $\dmf/h$                  &   $1.79(1)$ MHz/(V/cm)                   
                & Molecule-frame electric dipole moment         & This work \\
            $\wef/(2\pi)$                  &   $0.74(4)$ MHz                           
                & $\Omega$-doubling constant                    & \cite{Cossel:2012hna} \\
            $\wrot/(2\pi)$                  & $250$ kHz typ.
                & Rotation rate of $\Erot$                      & This work \\
            $g_F$                       &   $-0.0031(1)$                          
                & $F=3/2$ state g-factor                      & \cite{Loh:2013dc}\footnote{A sign error in Ref.~\cite{Loh:2013dc} has been corrected, however the magnitude of $g_F$ is unchanged.} \\
            $g_N$                   &   $5.25774(2)$                            
                & Nuclear magnetic g-factor of $^{19}$F       & \cite{Stone:2005ez}   \\
            $|\Eeff|/h$                 &   $5.63 \times 10^{24}$ Hz/($e\,$cm)  
                & Effective electric field & \cite{{Fleig:2013ij}} \\
        \hline \hline
        \end{tabular}
        \caption{Constants used in the effective Hamiltonian of Eq.~(\ref{eq:Heff}), for the calculation of frequency channels and systematic corrections. The total magnetic g-factor of $J=1, F=3/2$ states $g_F \equiv (\Gpar + g_N \mu_N/\mu_B)/3$ results from the combination of nuclear and electronic Zeeman effects.}
        \label{tab:constants}
    \end{table}

In our perturbative model of frequency channels, we treat the pairs of levels in either Stark doublet as a two-level system, and construct a $2\times2$ effective Hamiltonian for either doublet. We take the molecular rotation, Stark and Hyperfine Hamiltonian terms as an unperturbed Hamiltonian, then include the Zeeman, frame-rotation, $\Omega$-doubling, and eEDM terms as perturbations. While the Zeeman effect of $\Brot$ is the dominant diagonal contribution to the two-level effective Hamiltonian, perpendicular electric and magnetic fields, rotation, $\Omega$-doubling, and Stark mixing of rotational levels all contribute frequency shifts that are significant at the level of an eEDM search. In the basis of $F=3/2$, $m_F=\pm 3/2$ states, the two-level effective Hamiltonian has the form
    \begin{equation} \label{eq:Heff2}
        \Heff^{u/l} = \frac{1}{2} \begin{pmatrix}
        -3 (g_F \pm \delta g_F) \mu_B \Brot \pm 2 d_e |\Eeff| - 3 \alpha h \frot + \dots 
            & h (\Delta \pm \Delta^D) \\
        h (\Delta \pm \Delta^D)
            & 3 (g_F \pm \delta g_F) \mu_B \Brot \mp 2 d_e |\Eeff| + 3 \alpha h \frot + \dots
        \end{pmatrix},
    \end{equation}
where the $u/l$ superscript and upper or lower signs correspond to the upper or lower Stark doublet, $\alpha \equiv {\mathcal E}_Z/\Erot$ is the tilt of the electric field away from the $XY$ plane, and $\Delta$ is a rotation induced coupling with a Stark doublet-odd contribution $\Delta^D$. Here the magnetic g-factor $g_F$ results from the combined nuclear and electronic magnetic moments, which are coupled by the nuclear hyperfine Hamiltonian $\Hhf$. Notable features of this two-level system are the difference in magnetic g-factor between Stark doublets $\delta g_F$, the rotation-induced coupling $\Delta$, and the geometric frequency shift $3 \alpha \frot$. The magnetic g-factor difference arises from Stark mixing of $J=1$ with $J=2$ and from rotation-induced mixing of adjacent magnetic sublevels at second order in perturbation theory, and has the approximate form
    \begin{equation}
        \frac{\delta g_F}{g_F} \approx 
            -\frac{3 \dmf \Erot}{20 B_e} \left(1 - \frac{g_N \mu_N}{3 g_F \mu_B} \right)
            + \frac{3 \hbar^2 \wrot^2}{\dmf \Erot \Apar} \left(1 - \frac{2 g_N \mu_N}{3 g_F \mu_B} \right)
            \approx - 1 \times 10^{-3}.
    \end{equation}
The coupling $\Delta$ first arises at fourth order from the combined perturbations of $\Hrot$ and $\HOD$, and breaks the degeneracy of the $|m_F|=3/2$ states in either Stark doublet at $\Brot=0$ \cite{Leanhardt:2011jv,Meyer:2009fh}. The fourth-order perturbation theory expression for $\Delta$ and its Stark doublet-odd component $\Delta^D$ are given by
    \begin{equation}
        h \Delta = \frac{3 \hbar \wef}{2} \left( \frac{\hbar \wrot}{\dmf \Erot} \right)^3
            \left( \frac{18 \Apar^2 - 19 \dmf^2 \Erot^2}{\Apar^2 - \dmf^2 \Erot^2} \right), \qquad
        h \Delta^D = \frac{3 \hbar \wef}{2} \left( \frac{\hbar^3 \wrot^3}{\dmf^2 \Erot^2 \Apar} \right)
            \left( \frac{9 \Apar^2 - 8 \dmf^2 \Erot^2}{\Apar^2 - \dmf^2 \Erot^2} \right).
    \end{equation}
These expressions are valid as long as $\dmf\Erot \gg  \hbar \wef$ and $\dmf \Erot \gg  \hbar \wrot$. The strong scaling of $\Delta$ with $\Erot$ allows us to perform off-resonant $\pi/2$ pulses by modulating the magnitude of $\Erot$, as discussed in the main text. Finally, the geometric phase shift arises from the fact that if $\EZ$ is nonzero, the solid angle swept out by the rotating electric field differs from its nominal value of $2\pi$. This effect is discussed extensively in Ref.~\cite{Meyer:2006ky}. Higher order contributions to $\Heff^{u/l}$, denoted by ellipses in Eq.~(\ref{eq:Heff2}), are due to higher-order combinations of perturbing Hamiltonian components, and primarily result in small corrections to the terms already discussed. 

For our numerical studies, we use $\bm{\mathcal E}_{\rm ion}(t)$ and $\bm{\mathcal B}_{\rm ion}(t)$ resulting from the simulated ion motion (Section \ref{sec:ion_motion}) to construct $H_{\rm ion}(t) = H(\bm{\mathcal E}_{\rm ion}(t),\bm{\mathcal B}_{\rm ion}(t),\vecwrot)$, and numerically integrate the Schr\"odinger equation for a single ion by exponentiating $H_{\rm ion}(t)$ at each value of $t$ to obtain the time-evolution operator $U_{\rm ion}(t)$, which we apply to an initial state vector $\ket{\psi_0}$. For various sets of simulations we have either taken $\ket{\psi_0}$ to be one of states $\ket{a}$, $\ket{b}$, $\ket{c}$, or $\ket{d}$ [Fig.~2(a)] and included realistic $\pi/2$ pulses by modulating the simulated value of $\Erot$, or we have assumed an ideal $\pi/2$ pulse by initializing $\ket{\psi_0}$ as an equal superposition of states $\ket{a}$ and $\ket{b}$ or $\ket{c}$ and $\ket{d}$. Finally, we construct the (simulated) asymmetry ${\mathcal A}(t)$ by projecting $\ket{\psi(t)}$ onto the upper or lower doublet states, and fit ${\mathcal A}(t)$ using the functional form of Eq.~(1) in the main text. 

\subsection{Mixing of frequency channels} \label{sec:structure}
The non-negligible size of $\Delta$ and $\Delta^D$ throughout our spin precession experiment produces a unique structure of frequency channels that affects our systematics analysis, and warrants describing in more detail. To do so, we parametrize the two-state effective Hamiltonian in terms of ``diagonal'' and ``off-diagonal'' parity components $f_0^s$ and $\Delta^s$,
    \begin{equation}
        H_{\rm eff}(\tilS) = \frac{h}{2} \begin{pmatrix}
        f_0^0 + \tilB f_0^B + \tilD f_0^D + \dots 
            & \Delta + \tilB \Delta^B + \tilD \Delta^D \dots \\
        \Delta + \tilB \Delta^B + \tilD \Delta^D +  \dots 
            & f_0^0 + \tilB f_0^B + \tilD f_0^D + \dots
        \end{pmatrix}.
    \end{equation}
Expanding the frequency channels obtained from this Hamiltonian about $f_0^0$, we obtain
    \begin{align}
    f^0 &= |f_0^0| \left(1 
            + \frac{\Delta^2 + (\Delta^B)^2 + (\Delta^D)^2 + \dots}{2|f_0^0|^2} \right) 
        - f_0^B\frac{\Delta \Delta^B + \Delta^D \Delta^{BD} + \dots}{|f_0^0|^2} 
        - f_0^D\frac{\Delta \Delta^D + \Delta^B \Delta^{BD} + \dots}{|f_0^0|^2} + \dots \\
    f^B &= f_0^B \left(1 
            - \frac{\Delta^2 + (\Delta^B)^2 + (\Delta^D)^2 + \dots}{2|f_0^0|^2} \right) 
        + \frac{\Delta \Delta^B + \Delta^D \Delta^{BD} + \dots}{|f_0^0|} 
        - f_0^D \frac{\Delta \Delta^{BD} + \Delta^B \Delta^D + \dots}{|f_0^0|^2} + \dots \\
    f^D &= f_0^D \left(1 
            - \frac{\Delta^2 + (\Delta^B)^2 + (\Delta^D)^2 + \dots}{2|f_0^0|^2} \right) 
        + \frac{\Delta \Delta^D + \Delta^B \Delta^{BD} + \dots}{|f_0^0|} 
        - f_0^{B} \frac{\Delta \Delta^{BD} + \Delta^B \Delta^D + \dots}{|f_0^0|^2} + \dots \label{eq:fD_mixing} \\
    f^{BD} &= f_0^{BD} \left(1 
            - \frac{\Delta^2 + (\Delta^B)^2 + (\Delta^D)^2 + \dots}{2|f_0^0|^2} \right) 
        + \frac{\Delta \Delta^{BD} + \Delta^B \Delta^D + \dots}{|f_0^0|} 
        - f_0^{B} \frac{\Delta \Delta^D + \Delta^B \Delta ^{BD} + \dots}{|f_0^0|^2} + \dots
    \end{align}
with similar results for $f^{R}$, $f^{BR}$, $f^{DR}$ and $f^{BDR}$. Thus the nonzero value of $\Delta$, as well as any component of it that is odd under $B$, $D$, or $R$, causes mixing of ``diagonal'' parity components between measured frequency channels. This can cause systematic effects, the largest of which are described in Section~\ref{sec:systematics}. However, all $B$-odd components of $\Delta$ ($\Delta^B$, $\Delta^{BD}$ etc.) are negligible, greatly reducing the number of terms that must be considered. The regular form of these frequency channels also allows a straightforward correction that removes mixing terms up to third order in perturbation theory of $H_{\rm eff}$,
    \begin{equation} \label{eq:mixing_correction}
        f_0^{BD} = f^{BD}
            + f^B \left( \frac{f^D-f_0^D}{f^0} \right)
            + f^{BR} \left( \frac{f^{DR}-f_0^{DR}}{f^0} \right)
            + f^{BDR} \left( \frac{f^R-f_0^R}{f^0} \right) 
            + \text{h.\,o.}
    \end{equation}
The frequency channels $f^0$, $f^B$, $f^B$, $f^{BR}$, $f^{DR}$, $f^{BDR}$, and $f^R$ are measured simultaneously with $f^{BD}$, while the ``diagonal'' components $f_0^D$, $f_0^{DR}$ and $f_0^R$ must be estimated from theoretical models and auxiliary measurements. Note that in general the corrected value of $f_0^{BD}$ includes terms in addition to $2 d_e |\Eeff|$. The corrections in Eq.~(\ref{eq:mixing_correction}) account only for those systematics arising from the nonzero value of $\Delta$, and does not include ``diagonal'' systematics such as those arising from a difference in magnetic g-factor between Stark doublets and a non-reversing magnetic bias field. Under typical experimental conditions, the above corrections cancel mixing to the level of $\sim 10~\uHz$ ($\sim 10^{-30}~\ecm$). Correction terms are discussed in Section~\ref{sec:systematics} and are included in our uncertainty budget.

\subsection{Non-ideal frequency shifts} \label{sec:non_ideal}
A detailed description of every non-ideal frequency shift observed during our model-building phase is beyond the scope of this supplement, and will be given in a future publication. Here we list the parameters and experimental imperfections we explored in Table~\ref{tab:parameters}, and list observed effects and their observation channels in Table~\ref{tab:non_ideal}. More detailed descriptions of effects included in our uncertainty budget are given in Section \ref{sec:systematics}.
    \begin{table}[h]
        \centering
        \begin{tabular}{l l r}
        \hline \hline
            Description
                & Parameters 
                    & Study method \\ \hline
            Rotating ${\mathcal E}$ field
                & $\Erot$, $\wrot$ 
                    & E,M \\
            Axial magnetic bias gradient 
                & $\Baxgrad$, (${\mathcal B}_{2,m\neq 0}$, $\BX$, $\BY$, $\BZ$)\footnote{Here ${\mathcal B}_{2,m}$ indicates magnetic field gradients proportional to $\nabla [R^l Y_{lm}(\Theta,\Phi)]$ with $l=2$. In this notation, $\Baxgrad \equiv {\mathcal B}_{2,0}$.}
                    & E,M \\
            Non-reversing uniform ${\mathcal B}$ fields 
                & $\BXnr$, $\BYnr$, $\BZnr$
                    & E,M \\
            Non-reversing magnetic gradients 
                & ${\mathcal B}_{2,m\neq 0}^{\rm nr}$
                    & E,M \\
            Ion cloud position
                & $X_0$, $Y_0$, $Z_0$ 
                    & E,M \\
            Ion cloud secular motion amplitude 
                & $X_1$, $Y_1$, $Z_1$ 
                    & E,M \\
            Trap RF amplitude
                & $\Vrf$
                    & E,M \\
            Trap RF frequency
                & $\wrf$
                    & M \\
            Number of trapped \hff ions
                & $N$ 
                    & E \\
            $V_{\rm rot}$ harmonic distortion
                & $V_{nf}$, $\phi_{nf}$, $n=2\dots 6$ 
                    & E,M \\
            Currents in electrodes
                & 
                    & E,M \\
            Depletion laser polarization
                & 
                    & E \\
            Thermal drifts in trap amplifiers
                & 
                    & E,M \\
            Accumulating patch potentials
                & 
                    & E,M \\
        \hline \hline
        \end{tabular}
        \caption{Experimental parameters and imperfections explored as possible sources of systematic error. The letters E and M indicate that an effect was explored experimentally or through numerical and perturbative modeling, respectively.}
        \label{tab:parameters}
    \end{table}
    \begin{table}[h]
        \centering
        \begin{tabular}{l l l}
        \hline \hline
        Description
            & Channel
                & Scaling \\ \hline
        RF micromotion in $\Baxgrad$
            & $f^0$
                & $g_F \mu_B \Baxgrad (X_0^2 + Y_0^2) q /R_0$ \\
        Non-reversing $\Brot$
            & $f^B$
                & $g_F \mu_B \Baxgradnr$ \\
            & $f^{BD}$
                & $\delta g_{\rm eff} \mu_B \Baxgradnr $ \\
        $\Brotnr$ due to $\Erot$ 2nd harmonic, $\BXnr$, $\BYnr$
            & $f^B$
                & $g_F \mu_B V_{2f}(\BXnr X_0 + \BYnr Y_0)/\Erot R_0^2$ \\
            & $f^{BR}$
                & $g_F \mu_B V_{2f}(\BXnr Y_0 - \BYnr X_0)/\Erot R_0^2$ \\
        Stark-induced $D$-odd g-factor
            & $f^D$
                & $g_F \mu_B \Brot \dmf \Erot/B_e$ \\
        Rotation-induced $D$-odd g-factor
            & $f^D$
                & $g_F \mu_B \Brot (\hbar \wrot)^2/\dmf \Erot \Apar$ \\
        $\Delta$-induced effective $D$-odd g-factor
            & $f^D$
                & $\Delta^D \Delta / f^0$ \\
        Non-reversing axial ${\mathcal B}$-field
            & $f^R$
                & $(\Delta^2 / f^0)(g_F \mu_B \BZ/\hbar \wrot)$ \\
            & $f^{DR}$
                & $(\Delta^D \Delta / f^0)(g_F \mu_B \BZ/\hbar \wrot)$ \\
        $\Erot$ inhomogeneity-induced geometric phase
            & $f^{BR}$
                & $\frot (X_0^2+Y_0^2)Z_0 / R_0^3$ \\
        $\Brot$ due to electrode currents
            & $f^{BR}$
                & $g_F \mu_B \mu_0 \wrot C_{\rm elec} \Erot$ \\
        Axial secular motion frequency modulation
            & $f^{BR}$
                & $\beta Z_0 e \Erot / (m \wrot \omega_Z T) $ \\
        \hline \hline
        \end{tabular}
        \caption{Observed frequency shifts and their scaling with selected experimental parameters (numerical factors are omitted). Here $q$ is a dimensionless RF trap parameter \cite{Berkeland:1998cw}, $R_0$ is the radius of the RF trap, $Z_1$ is the amplitude of the center-of-mass secular motion of the ion cloud along the $Z$ axis, $C_{\rm elec}$ is the capacitance of an electrode, and $\beta \approx 1.4 \times 10^{-5}~{\rm mm}^{-2}$ is the fractional inhomogeneity of $\Erot$ along the $Z$ axis.}
        \label{tab:non_ideal}
    \end{table}
    %
    
\section{Systematics} \label{sec:systematics}
In perturbation theory of our effective Hamiltonian [Eq.~(\ref{eq:Heff})], an eEDM signal appears as the lowest-order contribution to the $f^{BD}$ frequency channel, and its value is independent of all experimental parameters. Any other contributions to this channel constitute systematic errors and must be corrected if they are large enough to cause a significant shift. Systematic shifts can generally be grouped into one of two categories: real frequency shifts arising from higher-order terms in the effective Hamiltonian (introduced in Section~\ref{sec:structure}), and apparent shifts arising from measurement errors. We have identified several possible sources of both types, and observed a frequency shift in $f^{BD}$ due to one effect.

We calculate systematic corrections according to a procedure similar to that of Ref.~\cite{Baron:2014ds}. Parametrizing a systematic shift in the eEDM channel as $f^{BD}_{{\rm syst},i} = P_i S_i$ where $P_i$ is a parameter and $S_i\equiv\partial f^{BD}/\partial P_i$, the corrected eEDM measurement in the $n^{\rm th}$ block is $f^{\rm eEDM}_n = f^{BD}_n + \sum_i f^{BD}_{{\rm corr},i,n} = f^{BD}_n - \sum_i P_{i,n} S_{i,n}$. We compute systematic corrections on a block-by-block basis, and we obtain the total $i^{\rm th}$ correction $\langle f^{BD}_{{\rm corr},i} \rangle$ and its uncertainty $\langle \delta f^{BD}_{{\rm corr},i} \rangle$ by propagating the standard errors $\delta P_{i,n}$ and $\delta S_{i,n}$ through the weighted averaging of blocks. For systematics that produced an observed shift in the eEDM channel (of which there was only one), we apply the correction and include $\langle \delta f^{BD}_{{\rm corr},i} \rangle$ in our uncertainty budget (Table~II). For systematics that we did not observe directly in the eEDM channel, we include a systematic uncertainty $\langle \delta f^{BD}_{{\rm tot},i} \rangle \equiv \sqrt{ \langle f_{{\rm corr},i}^{BD} \rangle^2 + \langle \delta f_{{\rm corr},i}^{BD} \rangle^2}$ in our uncertainty budget.

\subsection{Non-reversing $\Brot$ and effective differential g-factor} \label{sec:Brotnr}
\begin{figure}[h]
        \centering
        \includegraphics{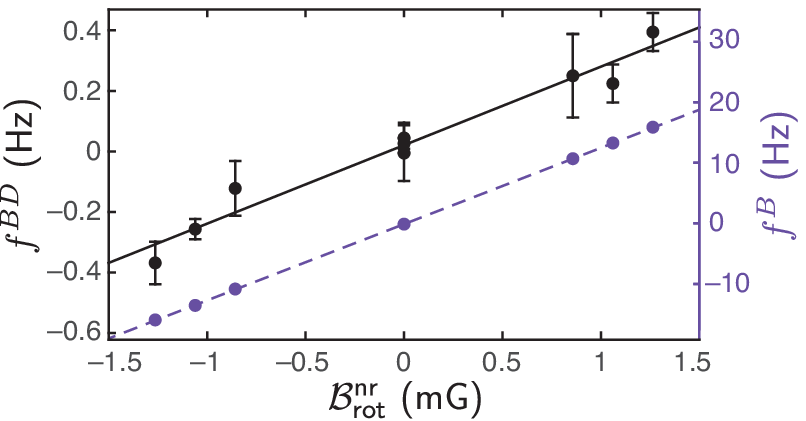}
        \caption{Systematic shift in the eEDM channel $f^{BD}$ due to non-reversing $\Brot$ (left hand vertical axis, solid line), and a proportional shift in the $f^B$ channel that we use to apply a correction (right hand vertical axis, dashed line). The constant of proportionality $f^{BD}/f^B = \delta g_{\rm eff}/g_F$ depends on $\Erot$ and $\frot$. The data shown here were taken under conditions chosen to make it particularly large ($\sim 10^{-2}$).}
        \label{fig:Brotnr_systematic}
    \end{figure}

The simplest contributor to a non-reversing rotating magnetic bias field $\Brotnr$ is an imperfect reversal of the applied axial gradient $\Baxgrad$. Contributions can also arise from other sources, including for example higher order magnetic gradients or time-dependent magnetic fields. Figure~\ref{fig:Brotnr_systematic} shows the only observed shift in the $f^{BD}$ channel, caused by a non-reversing axial gradient $\Baxgradnr$. Non-reversing ${\mathcal B}$ fields appear in the $f^{BD}$ channel due to the difference in the magnetic g-factors of the upper and lower doublet states $\delta g_F$, and due to the non-negligible size of $\Delta$ and $\Delta^D$ compared to the Zeeman shift. Fortunately, an amplified shift appears in the simultaneously collected $f^B$ channel, allowing us to apply a proportional correction. In terms of experimental parameters, the shift in the eEDM channel due to $\delta g_F$ and $\Delta^D$ is
    \begin{align} \label{eq:syst1}
        f^{BD}_{\rm syst,1} = 3 g_F \mu_B \Brotnr 
            \left( \frac{\delta g_F}{g_F} - \frac{\Delta \Delta^D}{|3 g_F \mu_B \Brot|^2} \right) + \text{h.\,o.}
    \end{align}
Under typical conditions, $f^{BD}_{\rm syst,1}$ is of order $\sim 10^{-3} f^B$ and the sum of higher order terms is $\sim 1~\uHz$. By comparison of the terms in Eq.~(\ref{eq:syst1}) to the leading order expressions for frequency channels in Table I in the main text, we find that we can apply a block-by-block correction (suppressing the subscript $n$)
    \begin{equation}
        f^{BD}_{\rm corr,1} = f^B \frac{\delta g_{\rm eff}}{g_F}  = f^B \left( \frac{f^D}{f^0} - 2\frac{\delta g_F}{g_F} \right).
    \end{equation}
The typical value of $f^B$ for a given block was $\lesssim 100~\mHz$, and due to our applied feedback to reduce the value of $|f^B|$, its average value over many consecutive blocks was much smaller. The small value of $\delta g_{\rm eff}/g_F$ makes the correction to $f^{BD}$ still smaller, with an average value of $\langle f^{BD}_{\rm corr,1} \rangle = \fcorrA(\sfcorrA)~\uHz$ over the entire eEDM dataset.
    %

\subsection{Geometric phase and axial magnetic field} \label{sec:geomphase}
While the six radial electrodes of our RF trap are designed to optimize uniformity of the rotating electric bias field, there nonetheless exist inhomogeneities in $\Erot$ that are well-described by spherical multipoles, as discussed in Section~\ref{sec:ion_motion}. The $(l,m) = (3,\pm1)$ and $(5, \pm 1)$ spherical multipole components of $\Erot$ cause a nonzero time-average tilt of the rotating electric field $\langle \alpha \rangle = \langle {\mathcal E}_Z/\Erot \rangle$ and a corresponding geometric frequency shift $f^{BR} \approx 3 \langle \alpha \rangle \frot$, shown in Fig.~3 of the main text. Independently, a non-reversing axial magnetic field $\BZnr$ produces a nonzero value of $\Delta^{DR}$, which forms the dominant contribution to the $f^{DR}$ channel. The mixing mechanism described in Section~\ref{sec:structure} leads to a shift in $f^{BD}$ equal to $-f_0^{BR} \Delta^{DR} \Delta/|f_0^0|^2$. Since $f^{BR} \approx f_0^{BR} = 3 \langle \alpha \rangle \frot$ and $f^{DR} \approx \Delta^{DR} \Delta/|f_0^0|$, we can in principle apply a corresponding correction
    \begin{equation} \label{eq:corr2}
        f^{BD}_{\rm corr,2} = \frac{f^{BR} f^{DR}}{f^0},
    \end{equation}
which has an average value of $\langle f^{BD}_{{\rm corr},2} \rangle = \fcorrB(\sfcorrB)~\uHz$ over the eEDM dataset. However, since this shift was too small for us to observe in the eEDM channel, we include a total systematic uncertainty $\langle \delta f^{BD}_{{\rm tot},2}\rangle = 4~\uHz$ in our uncertainty budget. 

Comparing Eqs.~(\ref{eq:mixing_correction}) and (\ref{eq:corr2}), we have neglected a contribution $-f^{BR} f_0^{DR}/|f_0^0|$ that cannot be obtained from our measured data channels. While $f_0^{DR}$ is typically negligible, a contribution to this ``diagonal'' channel can arise from an $R$-odd contribution to $\Erot$, and is discussed in Section~\ref{sec:Erot_R}. 

\subsection{Harmonic distortion of $\Erot$}
The systematic effects discussed in Sections~\ref{sec:Brotnr} and \ref{sec:geomphase} each arise from a single physical mechanism generating a frequency shift in $f^B$ and $f^{BR}$ respectively, which are subsequently ``mixed'' into $f^{BD}$ by the nonzero values of $\delta g_F$, $\Delta^D$, and $\Delta^{DR}$. Other physical effects that generate ``diagonal'' frequency shifts in $f^B$ and $f^{BR}$ enter into $f^{BD}$ in precisely the same way, and are thus contained in the corrections already applied. An illustrative example is the harmonic distortion of $\Erot$, which together with non-reversing uniform magnetic fields $\BXnr$ and $\BYnr$ produces a $\Brotnr$ $(\equiv \Brot^B)$ and a $\Brot^{BR}$ (an $R$-odd contribution to $\Brotnr$).

The rotating electric bias field $\Erot$ is generated by sinusoidal voltages of equal amplitude on each of the six radial electrodes, oscillating with a frequency of $\frot$ and with a relative phase of $\pi/3$ between adjacent electrodes. These voltages are generated by power operational amplifiers, which inevitably suffer from harmonic distortion. While exploring sources of systematic error, we observed frequency shifts up to several Hz in the $f^B$ and $f^{BR}$ channels. The observed shifts had a linear dependence on both transverse uniform magnetic fields and the equilibrium position of the ion cloud during the spin precession experiment. These shifts, shown in Fig.~2 of the main text, were caused by a contribution to $\Brotnr$ from the combined effect of transverse magnetic fields and an oscillating electric field gradient generated by the second harmonic of $\Erot$. From a simple model of electric fields in the ion trap and using 2nd order perturbation theory of our effective Hamiltonian, we obtained model expressions that matched the observed frequency shifts,
    \begin{equation} \label{eq:fB_fBR_harmonic}
        f^B = - \frac{3 g_F \mu_B V_{2f}}{4 \Erot R^2} 
            (\BX X_0 + \BY Y_0) \cos{\phi_{2f}}, \qquad
        f^{BR} = - \frac{3 g_F \mu_B V_{2f}}{4 \Erot R^2} 
            (\BX Y_0 - \BY X_0) \sin{\phi_{2f}},
    \end{equation}
where $V_{2f}$ and $\phi_{2f}$ are the amplitude and phase of the 2nd harmonic. During eEDM data collection, we suppressed $V_{2f}$ to $-70$~dBc by adding a feedforward signal to the voltages generating $\Erot$, and canceled $\BX$ and $\BY$ to within 30~mG of zero at the RF trap center using magnet coils. Both of these frequency shifts can cause a false eEDM; $f^B$ through the effective differential g-factor, and $f^{BR}$ through a shift in the $f^{DR}$ channel. Both, however, are already corrected by $f^{BD}_{\rm corr,1}$ and $f^{BD}_{\rm corr,2}$. Higher harmonics of $\frot$ combined with magnetic gradients can contribute higher-order terms to Eq.~(\ref{eq:fB_fBR_harmonic}), however the resulting shifts in the $f^{BD}$ channel are similarly accounted for by the corrections already applied.

\subsection{Frequency modulation due to axial secular motion}
\begin{figure}[ht]
        \centering
        \includegraphics{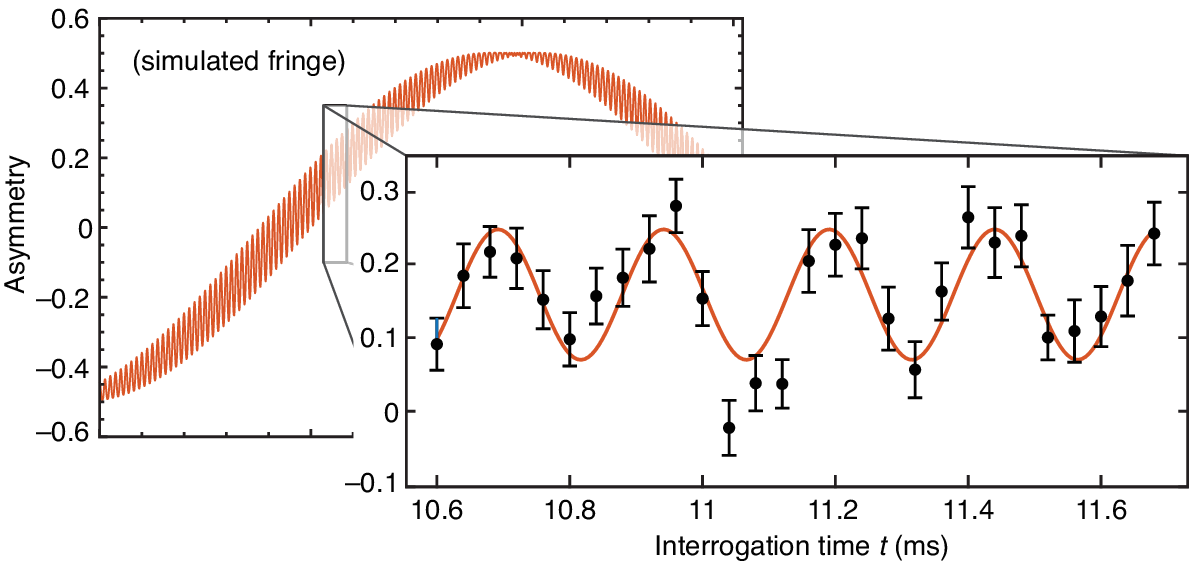}
        \caption{Phase modulation of an interference fringe at $f_Z = 4.01(7)~\kHz$ due to axial secular motion of the ion cloud with an intentionally exaggerated amplitude of $\sim 3~{\rm mm}$, measured near the steepest portion of the fringe where $2\pi f t \approx \pi/2$.}
        \label{fig:freq_mod}
    \end{figure}

The geometric frequency shift $3 \alpha \frot$ in Eq.~(\ref{eq:Heff2}) generates a $BR$-odd frequency contribution proportional to an axial electric field ${\mathcal E}_Z$, as was already discussed in the context of geometric phases. While coherent axial secular motion of the ion cloud at frequency $f_Z \gg f^0$ will not produce a time-average nonzero value of $\alpha$, it does cause a $BR$-odd modulation of the instantaneous spin precession frequency, as shown for a deliberately large secular motion amplitude in Fig.~\ref{fig:freq_mod}. The ponderomotive potential associated with $\Erot$ inhomogeneity provides a source for axial secular motion, since our $\pi/2$ pulses involve modulating $\Erot$ and thus applying position-dependent impulses to the ion cloud. Further, we apply $\pi/2$ pulses of different lengths in the upper or lower Stark doublet, which can lead to a $D$-odd slosh amplitude and therefore a $BDR$-odd modulation. 

We perform a frequency measurement by collecting a set of (typically) six equally spaced points at short interrogation time $t\lesssim 50~\ms$ and a second set at late time $t\lesssim 700~\ms$, each set spanning approximately one period of spin precession. The frequency resulting from a nonlinear least squares fit to the functional form of Eq.~(1) is approximately equal to the difference in phase between early and late time, $f \simeq (\phi_T - \phi_0)/2\pi T$. If aliasing this frequency modulation causes a systematic phase shift, a systematic frequency shift $f^{BR} = \Delta \phi^{BR}/2\pi T$ will result (with a similar frequency shift due to $\phi^{BDR}$). To the extent that the spacing of our asymmetry points in interrogation time is incommensurate with $1/f_Z$, the shift in $f^{BR}$ and $f^{BDR}$ will be suppressed. However, we did not vary either $f_Z$ or the spacing of our interrogation times during eEDM data collection in order to enhance this suppression. Thus a contribution to offsets in $f^{BR}$ and $f^{BDR}$ from this effect may be present. 

While any $BR$-odd contribution from frequency modulation is a ``diagonal'' frequency shift and is accounted for by $f^{BD}_{\rm corr,2}$, the $BDR$-odd frequency shift has not yet been accounted for. This is done so in the same manner as $f^{BD}_{\rm corr,1}$ and $f^{BD}_{\rm corr,2}$, as outlined in Section~\ref{sec:structure}. The corresponding correction is
    \begin{equation}
        f^{BD}_{\rm corr,3} = \frac{f^{BDR} f^R}{f^0}.
    \end{equation}
Since we did not observe a shift in the eEDM channel due to this systematic, we include an overall uncertainty $\langle \delta f^{BD}_{{\rm tot},3} \rangle = \sftotC~\uHz$. As in the case of $f^{BD}_{\rm corr,2}$ where we neglected $f_0^{DR}$, here we have neglected $f_0^R$. The main known source of $f_0^R$ is discussed in the next section.

\subsection{Rotation-odd $\Erot$} \label{sec:Erot_R}
As discussed in Section~\ref{sec:data_collection}, the $R$ switch is controlled digitally by adjusting the relative phases of six DDS channels, each of which provides the $\Erot$ signal for one electrode. As a result, we expect an $R$-odd rotating electric field $\Erot^R$ to be very small, possibly dominated by RF pickup between neighboring amplifier circuits in our ion trap driver electronics. To the extent that $\Erot^R$ does exist, it could potentially cause a systematic error through accidental cancellation of the $\BZnr$-induced $f^R$ and $f^{DR}$ described in Section~\ref{sec:geomphase}. This could occur because $\Erot^R$ produces $R$- and $DR$-odd ``diagonal'' frequency components,
    \begin{equation}
        f_0^R = 3 g_F \mu_B \Brot^R, \qquad
        f_0^{DR} = 3 \delta g_{\rm eff} \mu_B \Brot^R,
    \end{equation}
while the shifts in the same channels from $\BZnr$ arise from $\Delta^R$ and $\Delta^{DR}$. The two sources of $f^R$ and $f^{DR}$ cannot be distinguished, and lead to systematic shifts of opposite sign in the eEDM channel. However, at our present level of sensitivity, all of these shifts were small compared to our statistical uncertainty. The resulting systematic frequency shifts in $f^{BD}$ due to $\Erot^R$ are
    \begin{equation}
        f^{BD}_{\rm syst,4} = f^{BR} \frac{\delta g_{\rm eff}}{g_F} \frac{\Erot^R}{\Erot},\qquad 
        f^{BD}_{\rm syst,5} = f^{BDR} \frac{\Erot^R}{\Erot}.
    \end{equation}
Since we did not monitor $\Erot^R$ throughout data collection, we use a very conservative estimate of $\Erot^R/\Erot = 0.01$ with an uncertainty of $\delta \Erot^R/\Erot = 0.01$, and calculate systematic uncertainties $\langle \delta f^{BD}_{\rm tot,4} \rangle$ and $\langle \delta f^{BD}_{\rm tot,5} \rangle$. The very small value of $\delta g_{\rm eff}/g_F \approx 10^{-3}$ suppresses $f^{BD}_{\rm tot,4}$ to $\sim 1~\uHz$, so we include only $\langle \delta f^{BD}_{\rm tot,5} \rangle = \sftotE~\uHz$ in our uncertainty budget.

\subsection{Doublet population contamination}
\begin{figure}[h]
        \centering
        \includegraphics{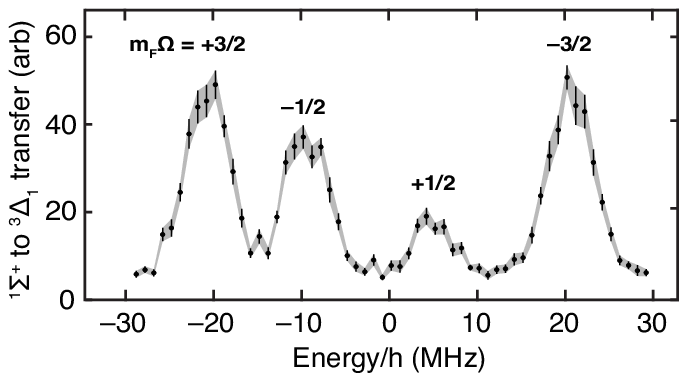}
        \caption{Stimulated Raman transfer from $^1\Sigma^+$, $J=0$ to $\td$, $J=1$, $F=3/2$, showing Doppler-broadened resonances at the locations of Stark doublets.}
        \label{fig:scan_toptica}
    \end{figure}

\begin{figure}[h]
        \centering
        \includegraphics{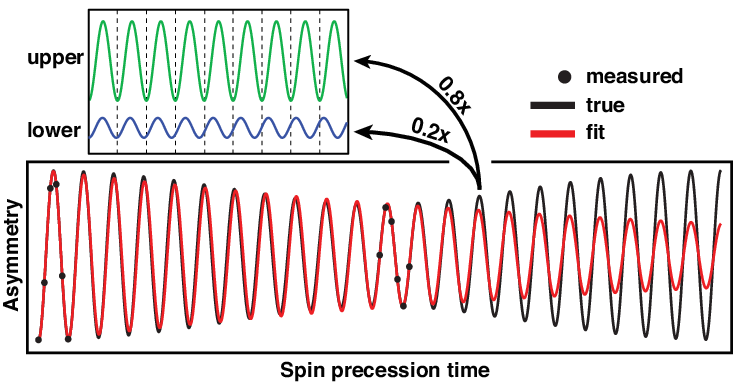}
        \caption{Simulated effect of doublet population contamination with a fraction $w=0.2$ of the $\td$ population in the undesired Stark doublet. Green and blue traces in the inset illustrate the fractional contributions from the upper and lower doublet to the total true interference fringe (black line). Measuring the asymmetry at only short and long times (black points) leads to a misinterpretation of beating as a loss of coherence (red line), and results in an error in the fit frequency.}
        \label{fig:beating}
    \end{figure}

When population is transferred from $^1\Sigma^+$ to $^3\Delta_1$, the detuning of the second transfer laser is set by an acousto-optic modulator to select either the upper or lower Stark doublet to be populated. The upper and lower Stark doublets are resolved by approximately nine times the $1\sigma$ Doppler width of each transfer resonance (Fig.~\ref{fig:scan_toptica}), so population of the undesired Stark doublet is highly suppressed. During both population transfer and strobed depletion, however, spontaneous decay from the $^3\Pi_{0^+}$ and $^3\Sigma_{0^+}^-$ excited states to all hyperfine levels in $\td$, $J=1$ can occur, albeit with a very small probability. Because our depletion and dissociation state readout processes are not Stark doublet-selective, population in the undesired Stark doublet will lead to uncharacterized beating in our interference fringes that will be misidentified as a loss of coherence and a frequency shift, as shown in Fig.~\ref{fig:beating}. The effect of the apparent frequency shift is to suppress the measured value of $f^D$, leading to two sources of systematic error that we have identified. 

The first systematic error arising from population in the wrong Stark doublet is that a suppressed value of $f^D$ will suppress the correction of Eq.~(\ref{eq:syst1}). Including this effect, we predict a value for $f^D$ of
    \begin{equation}
        f^D_{\rm meas} = f^D_{\rm true} - \frac{w}{2 \pi T} \sin{(4 \pi f^D_{\rm true} T)},
    \end{equation}
where $T$ is the temporal separation between early and late time fringe measurements, and $w$ is the fraction of the $\td$ state population in the lower (upper) Stark doublet when only the upper (lower) doublet should be populated. The resulting systematic error in the correction (\ref{eq:syst1}) is
    \begin{equation}
        f^{BD}_{\rm syst,6} = \frac{w f^B}{2\pi f^0 T} \sin{(4 \pi f^D_{\rm true} T)}.
    \end{equation}
Using this model and the difference between our measured and predicted values of $f^D$, we estimate $w = 0.02(1)$. However, $w$ was not directly monitored during our data collection and systematic errors in the measured parameters contributing to the predicted value of $f^D$ may dominate over any true population contamination. We estimate the total systematic error due to this effect to be $\langle \delta f^{BD}_{{\rm tot},6} \rangle = \sftotF~\uHz$.

The second systematic error introduced by population contamination is proportional to a phase shift $\phi^{BD}$ that has the same parity as the eEDM frequency channel. While in neutral beam experiments, an uncharacterized $\phi^{BD}$ shift leads directly to a systematic frequency shift $\phi^{BD}/2 \pi T$, our practice of measuring both early and late time phase nominally distinguishes $\phi^{BD}$ from $f^{BD}$. In the presence of population contamination, a systematic shift will be present and of the form
    \begin{equation}
        f^{BD}_{\rm syst,7} = \frac{w}{ \pi T} \sin{(2 \phi^{BD})} \sin^2{(2\pi f^D T)}.
    \end{equation}
Again, since $w$ was not directly monitored during data collection, and since inconsistencies in our measured and predicted values of $f^D$ could be due to other sources, we include a total systematic uncertainty $\langle f^{BD}_{{\rm tot},7} \rangle = \sftotG~\uHz$ in our uncertainty budget. This contribution dominates over $f^{BD}_{{\rm tot},6}$, and is the largest contributor to systematic error in our experiment. Improved monitoring of $w$ in the next generation of this experiment (via, e.g., microwave spectroscopy of the $\td$ $J=1 \rightarrow J=2$ transition), as well as operating in a regime of $\Erot$ and $\frot$ where $f^D$ is suppressed, will reduce this systematic to the order of $\sim1~\uHz$.

\bibliographystyle{apsrev4-1} 
\bibliography{supplement.bbl}

%% file: macros.tex
\newcommand{\dgeff}{{\delta g}_{\rm eff}}
\newcommand{\dmf}{d_{{\rm mf}}}

\newcommand{\Eeff}{{\mathcal E}_{\rm eff}}

\newcommand{\Erot}{{\mathcal E}_{\rm rot}}

\newcommand{\vecErot}{\bm{\mathcal E}_{\rm rot}}

\newcommand{\Brot}{{\mathcal B}_{\rm rot}}

\newcommand{\Baxgrad}{{\mathcal B}'_{\rm axgrad}}

\newcommand{\Brotnr}{{\mathcal B}_{\rm rot}^{\rm nr}}

\newcommand{\BYnr}{{\mathcal B}_Y^{\rm nr}}

\newcommand{\frot}{f_{\rm rot}}

\newcommand{\frf}{f_{\rm rf}}

\newcommand{\td}{{^3\Delta_1}}

\newcommand{\sgn}{{\rm sgn}}

\newcommand{\tilB}{{\tilde{B}}}
\newcommand{\tilD}{{\tilde{D}}}
\newcommand{\tilR}{{\tilde{R}}}

\newcommand{\tilS}{{\tilde{S}}}

\newcommand{\ecm}{e\,{\rm cm}}

\newcommand{\Hz}{{\rm Hz}}
\newcommand{\uHz}{\mu{\rm Hz}}

\newcommand{\MHz}{{\rm MHz}}
\newcommand{\mHz}{{\rm mHz}}



%% file: results_macros.tex
\newcommand{\fcorrA}{-1}
\newcommand{\sfcorrA}{5}

\newcommand{\sftotB}{4}

\newcommand{\sftotC}{2}

\newcommand{\sftotE}{14}

\newcommand{\sftotG}{195}
\newcommand{\feedmmHz}{0.10}
\newcommand{\sstatmHz}{0.87}
\newcommand{\ssystmHz}{0.20}
\newcommand{\sstatuHz}{868}
\newcommand{\ssystuHz}{195}
\newcommand{\stotuHz}{890}
\newcommand{\deecm}{0.9}
\newcommand{\sstatecm}{7.7}
\newcommand{\ssystecm}{1.7}
\newcommand{\ubecm}{1.3}
\newcommand{\mhztoecm}{1.13}